\documentclass[12pt,aps,nofootinbib]{revtex4}
\usepackage{amsmath}
\usepackage{graphicx}
\DeclareMathOperator{\erf}{erf}

\def\gsim{\gtrsim}
\def\lsim{\lesssim}

\newcommand{\Gammamat}{\boldsymbol \Gamma}
\newcommand{\Rmat}{\mathbf R}
\renewcommand{\vec}[1]{\boldsymbol #1}

\DeclareMathOperator{\Trace}{Tr}
\DeclareMathOperator{\diag}{diag}
\DeclareMathOperator{\sign}{sgn}
\newcommand{\dtn}[1]{\frac{d^3{\vec #1}}{(2\pi)^3}\,}
\begin{document}

\title{Pairing in Many-Fermion Systems:\\ An Exact Renormalisation Group Treatment}
\author{Michael C. Birse$^1$}
\author{Boris Krippa$^{1,2}$}
\author{Judith A. McGovern$^1$}
\author{Niels R. Walet$^2$}
\affiliation{$^1$Theoretical Physics Group, Department of Physics and Astronomy,
University of Manchester, M13 9PL, UK}
\affiliation{$^2$Department of Physics, UMIST,
P.O. Box 88, Manchester, M60 1QD, UK}
\date{\today}
\begin{abstract}
We study the application of the exact renormalisation group to a 
many-fermion system with a short-range attractive force. We introduce a 
boson field to describe pairing effects, and take a simple ansatz for 
the effective action. We derive a set of approximate flow equations for 
the effective coupling including boson and fermionic fluctuations. The 
initial conditions are obtained by renormalising the interaction to fit 
the scattering length in vacuum. At some critical value of the running 
scale, the numerical solutions show a phase transition to a gapped 
phase. Standard results are recovered 
if we omit the boson loops. When boson fluctuations are 
included, we find that their contributions are significant only in the 
small-gap regime.

\end{abstract}
\maketitle

Attractive forces between fermions, even if they are too weak to
produce two-body bound states, play a crucial role in many areas of
many-body physics. In a system of fermions, any attraction, 
no matter how weak, can cause such particles to form correlated 
``Cooper pairs'' at zero temperature.
Examples range from superfluidity in liquid helium-3 to colour 
superconductivity in dense quark matter. The ground state of the
system becomes qualitatively different through this pairing,
as can be seen from the occurrence of a phase transition at
some critical temperature. The order parameter for this is the
energy gap which appears in the fermion spectrum. Within such
systems we can identify two extreme limits, depending on
the interaction strength. The weak-coupling regime, with no
two-body bound states, manifests itself in
Bardeen-Cooper-Schrieffer (BCS) superconductivity, while the
strong-coupling regime, with a deeply bound state which
approximates an elementary boson, corresponds to Bose-Einstein
Condensation (BEC).

In this letter we study pairing within a framework inspired by modern
effective field theory (EFT). An EFT description of phenomena is
intended to be generic, independent of details of the
underlying theory, and depending only on the degrees of freedom and
interactions relevant at the energy scale being considered
\cite{BvK02}. An interesting, but rather ad-hoc approach to
combining pairing with EFT ideas is the work of Papenbrock and
Bertsch \cite{PB99}. A more detailed analysis of the same problem, but
cast in a slightly different language, can be found in the condensed
matter literature \cite{MPS98}, and has a long history
\cite{Tradition} (see also \cite{Babaev}). There is also some 
closely related work by Weinberg \cite{wein94}, concentrating on the 
effective potential in a field-theoretic framework.

Papenbrock and Bertsch \cite{PB99} use a standard BCS approach with a 
contact interaction, which they relate to scattering observables by
a subtractive renormalisation scheme. Their results are limited to 
positive chemical potential but many of their basic equations
are identical to those of Marani {\em et al.}~\cite{MPS98}. In the 
latter work the chemical potential $\mu$ is not tied to the Fermi 
momentum $p_F$, making it possible to study the crossover to BEC, 
which occurs for negative values of $\mu$.
In order to find a more complete way to include many-body physics
beyond the mean field, within an EFT-inspired approach, we have used a
method that draws together ideas from Ref.~\cite{wein94} and 
Refs.~\cite{PB99,MPS98}. The approach is based on the use of the Exact 
Renormalisation Group (ERG) \cite{Wi74,Wett93,BTW02}.

The goal of the ERG approach is to construct the Legendre transform of
the effective action: $\Gamma[\phi_c]=W[J]-J\cdot \phi_c$, where $W$
is the usual partition function in the presence of an external source
$J$ \cite{wein96}. The action functional $\Gamma$ generates the 
1PI Green's functions for small fluctuations around the ground
state, and it reduces to the effective potential for homogeneous 
systems. Instead of trying to evaluate it directly from $W[J]$, one
can use an artificial renormalisation group flow, created by introducing 
an artificial gap in the energy spectrum for the fields.
This depends on a momentum scale $k$ since we can define the
effective action by integrating over components of the 
fields with $q \gsim k$. The RG trajectory then interpolates between the 
classical action of the underlying field theory (at large $k$), and the 
full effective action (at $k=0$) \cite{Wett93,BTW02}. This method has
been successfully applied to a range of problems, from condensed matter
physics \cite{TW94,DMT03} to particle physics \cite{JW96,BJW00}.

Here we study a system of fermions interacting through an attractive 
two-body contact potential.  We take as our starting point an EFT that 
describes the $s$-wave scattering of two fermions with a $T$-matrix 
determined by the scattering length $a_0$: $T=-(4\pi/M)(-1/a_0-i p)^{-1}$.
Here a positive scattering length corresponds to a system with a two-body 
bound state (and hence repulsive phase-shifts for low-energy scattering) 
and a negative scattering length to one without a bound state.
The binding energy gets deeper as $a_0$ gets smaller,
while the limit $a_0\rightarrow\pm\infty$ corresponds to a zero-energy 
bound state.

Since we are interested in the appearance of a gap in the fermion
spectrum, we need to parametrise our effective action in a way that
can describe the qualitative change in the physics when this occurs.
A natural way to do this is to introduce a boson field whose vacuum 
expectation value (VEV) describes that gap and so acts as the 
corresponding order parameter \cite{wein94}. This ERG approach is similar 
to that for the chiral phase transition in Refs.~\cite{JW96,BJW00}. At  
the start of the RG flow, the boson 
field is not dynamical and is introduced through a 
Hubbard-Stratonovich transformation of the four-point interaction.
As we integrate out more and more of the fermion degrees of freedom by 
running $k$ to lower values, we generate dynamical terms in the bosonic
effective action. 

For simplicity we treat a single species of fermion, as in neutron
matter. We introduce a boson field $\phi$ describing correlated
pairs of fermions, which leads rather naturally to the following
Ansatz for $\Gamma$:
\begin{eqnarray}
\Gamma[\psi,\psi^\dagger,\phi,\phi^\dagger,\mu,k]&=&\int d^4x\,
\left[\phi^\dagger(x)\left(Z_\phi\, i \partial_t 
+\frac{Z_m}{2m}\,\nabla^2\right)\phi(x)-U(\phi,\phi^\dagger)\nonumber\right.\\
&&\qquad\qquad+\psi^\dagger\left( Z_\psi (i \partial_t+\mu)
+\frac{Z_M}{2M}\,\nabla^2\right)\psi\nonumber\\
&&\qquad\qquad\left.- Z_g\left(\frac{i}{2}\,\psi^{\rm T}\sigma_2\psi\phi^\dagger
-\frac{i}{2}\,\psi^\dagger\sigma_2\psi^{\dagger{\rm T}}\phi\right)\right],
\label{eq:Gansatz}
\end{eqnarray}
The field $\phi$ has a non-standard normalisation, since 
we shall relate its VEV to the gap $\Delta$ via $\Delta^2=\langle
\phi^\dagger\phi\rangle$. This means that we have only a dimensionless
coupling-constant renormalisation $Z_g$ for the boson-fermion
coupling. The bosons carry twice the charge of a fermion, and so couple to 
the chemical potential via a term $2\mu Z_\phi\phi^\dagger\phi$ which has 
been absorbed into the definition of the quadratic term in the potential $U$. 
(Here $M$ is the mass of the fermions in vacuum and $m$ is naively
chosen to have the value $2M$, but its real role is only to make $Z_m$
dimensionless.) 

We expand the potential $U$ about its minimum to quartic
order in the field \cite{BTW02,DMT03},
\begin{equation}
U(\phi,\phi^\dagger)= u_0+ u_1(\phi^\dagger\phi-\Delta^2)
+\frac{1}{2}\, u_2(\phi^\dagger\phi-\Delta^2)^2,
\label{eq:potexp}
\end{equation}
where the $u_n$ are defined as the derivatives of $U$ at its minimum, which 
occurs where $\phi^\dagger=\phi=\Delta$. The values of the $u_n$ run with
the scale of the regulator, $k$. They also depend on the chemical
potential $\mu$, the other external parameter in $\Gamma$.

In the RG evolution, we start at high $k$ from a free bosonic action
and gapless fermions. In this symmetric phase we have $u_1>0$
and $\Delta=0$, and so the VEV of $\phi$ is zero. When $k$ is lowered, 
we expect $u_1$ to decrease until it reaches zero.  At this point there 
is a transition to a phase with spontaneously broken $U(1)$ symmetry 
and a fermion energy gap. Here the minimum moves away from $\Delta=0$ and, 
since we expand around the minimum of $U$, we take $u_1=0$ in 
Eq.~(\ref{eq:potexp}). The bosonic excitations in this phase are gapless 
``Goldstone'' bosons.

The other parameters in the action, namely the wave-function renormalisations 
$Z_{\phi,\psi}$, the kinetic mass renormalisations $Z_{m,M}$ and the coupling
$Z_g$, are all evaluated in the background field corresponding to the minimum 
of the potential.
These other parameters also run with $k$. For large values of $k$, the 
fermionic parameters $Z_M$ and $Z_\psi$ tend to unity, while the bosonic 
ones $Z_\phi$ and $Z_m$ tend to zero. We choose the normalisation of the 
field $\phi$ so that $Z_g$ tends to unity for large $k$.
The chemical potential $\mu$ can be determined from 
$\partial \Gamma/\partial \mu=n$, where $n$ is the baryon-number density. 
To be able to study the whole range of phenomena from BCS to BEC 
(where $\mu$ becomes negative) we keep this density fixed, and 
allow $\mu$ to run during the RG evolution. Following 
Refs.~\cite{MPS98,Babaev} we define a ``Fermi momentum'' $p_F$ in terms of
density: $p_F=(3 \pi^2 n)^{1/3}$. In the symmetric phase the fermion spectrum 
does not change and 
$p_F$ can be related to $\mu$ by $\mu=p_F^2/(2M)$. However once a gap appears
this connection is lost and indeed the whole idea of a Fermi surface may 
lose its meaning. 

In the results presented here, we allow only $Z_\phi$ and the parameters in
the potential to run independently since this is the minimal set needed
to treat the bosons dynamically. We freeze the other coefficents 
($Z_\psi$, $Z_M$ and $Z_g$) to unity or, in the case of the boson kinetic mass, 
we have also explored setting $Z_m=Z_\phi$. 

The evolution equation for $\Gamma$ in the ERG has a straightforward
one-loop structure \cite{BTW02}. For constant $\mu$ it can be written 
\begin{equation}
\partial_k\Gamma=-\frac{i}{2}\,\Trace \left[
(\Gammamat ^{(2)}_{BB}-\Rmat_B)^{-1}\,\partial_k\Rmat_B\right]
+\frac{i}{2}\,\Trace \left[
(\Gammamat ^{(2)}_{FF}-\Rmat_F)^{-1}\,\partial_k\Rmat_F\right].
\label{eq:Gamevol}
\end{equation}
Here $\Gammamat ^{(2)}_{FF(BB)}$ is the matrix containing second
functional derivatives of the effective action with respect to the
fermion (boson) fields and $\Rmat_{B(F)}$ is a matrix containing the
corresponding boson (fermion) regulators. A $2\times 2$ matrix structure 
arises for the bosons because we treat $\phi$ and $\phi^\dagger$ as
independent fields in order to include the number-violating condensate.
A similar structure also appears for the fermions. By inserting our
ansatz for $\Gamma$ into this equation we can turn it into a set of
coupled equations for the various parameters discussed above.

The one-loop structure of the ERG equations means that the evolution at
large $k$ can be matched onto that of the underlying EFT \cite{BvK02},
albeit for a non-standard choice of regularisation.
In the bosonic sector, the regulator is an additional
bilinear term in the fields. It has the matrix structure 
\begin{equation}
\Rmat_B(q,k)=R_B(q,k)\,\diag(1,1),
\end{equation}
where $R_B(q,k)$ is a scalar function. This regulator provides an extra 
contribution to the single-particle energies, which should suppress the 
contributions of states with momenta $q\lsim k$. As $k$ tends to zero 
$R_B(k)$ should vanish, so that we would recover the full effective 
action in the absence of any truncations.
For large $k$, $R_B(q,k)$ should be large for momenta $q\lsim k$, to give
all these modes large energies. In particular, it should be of order
$k^2$ if the behaviour of any integral for large $k$ is to reflect its
order of divergence. For more discussion of the choice of regulator in this
approach, see Ref.~\cite{Litim}.

In the symmetric phase, our regulator for the fermions should be positive for 
particle states ($q>p_F$) and negative for hole states ($q<p_F$), so that it
provides a $k$-dependent energy gap for excitations around the Fermi surface. 
It should have the structure
\begin{equation}
\Rmat_F(q,p_F,k)=\sign\bigl(\epsilon(q)-\mu\bigr)\,R_F(q,p_F,k)\,\diag(1,-1),
\end{equation}
where $\epsilon(q)=q^2/(2M)$.
The function $R_F(q,p_F,k)$ should be peaked about $p_F$ to suppress the 
contributions of states with momenta $|q-p_F|\lsim k$, 
but otherwise it should behave like the bosonic function.
Note that the ``Fermi surface'' in the gapped phase
no longer lies exactly at $p_F$, and there may not even be a well-defined 
Fermi surface. However, in that case a real gap has appeared in the spectrum 
and the regulator no longer plays a crucial role.

It is worth noting that we can recover the usual mean-field results if we 
include fermion loops only, ignoring all diagrams with virtual bosons. 
In this case the equation for the effective potential can be integrated 
analytically without making any truncation, and the results for $k=0$ coincide 
with those in Refs.~\cite{PB99,MPS98}. The initial conditions on the evolution 
are obtained by assuming that in vacuum our theory reproduces the scattering 
length $a_0$, and that the necessary subtraction are identical in matter and
in vacuum. (See Ref.~\cite{BvK02} and references therein for discussion
of the vacuum problem.) The resulting running potential has the form
\begin{equation}
U^{\text{MF}}(\Delta^2,\mu,k)=\int \dtn{q}
\left(\epsilon(q)-\mu+\frac{1}{2}\frac{\Delta^2}{\epsilon(q)}
-\sqrt{E_{FR}(q,k)^2+\Delta^2}\right
)+\frac{M}{4\pi
a_0} \Delta^2,\label{eq:UF}
\end{equation}
where we have introduced the short-hand notation
\begin{equation}
E_{FR}(q,k)=\epsilon(q)-\mu+R_F(q,p_F,k)\,
\sign\bigl(\epsilon(q)-\mu\bigr).
\end{equation}

For $k=0$ this has a closed-form expression in terms of an associated
Legendre function, $P_l^m(y)$:
\begin{equation}
U^{\text{MF}}(\Delta^2,\mu,0)=\frac{{ {k_\Delta }}^5}{2 M\pi} 
\left( \frac{1}{8  {a_0}  {k_\Delta } } - \frac{1}{15  }
{\left( 1 +  { {x_0}}^2 \right) }^{\frac{3}{4}} 
P_{\frac{3}{2}}^1\left(- \frac{ {x_0}}{{\sqrt{1 + { {x_0}}^2}}}
\right) \right).
\end{equation}
where $k_\Delta=\sqrt{2M\Delta}$ and $x_0=\mu/\Delta$.
In the weakly attractive limit, $p_Fa_0\rightarrow 0^-$, the chemical
potential is directly related to the density by $\mu=\epsilon_F=p_F^2/(2M)$,
and minimising $U^{\text{MF}}$ with respect to $\Delta$ leads to the standard 
result (see, e.g., Ref.~\cite{PB99})
\begin{equation}
\Delta=\frac{8}{e^2}\, \epsilon_F \exp\left(-\frac{\pi}{2 p_F |a_0|} \right).
\label{eq:lowd}
\end{equation}

More generally, working at constant density, we can minimise $U$ subject to 
the subsidiary condition $\partial U/\partial\mu=n$. This gives the results from 
Ref.~\cite{MPS98}, which can be turned into a pair of equations for $\Delta$ 
and $\mu$,
\begin{eqnarray}
\Delta/\epsilon_F&=&\bigl(-3\pi^2 (x_0^2+1)^{1/4} 
P^1_{1/2}(y_0)\bigr)^{-2/3}\nonumber\\
\noalign{\vskip 5pt}
\frac{1}{p_Fa_0}&=&-\frac{2}{3}\, x_0(x_0^2+1)^{1/4} \left(
P^1_{1/2}\left(y_0\right)-\frac{1}{y_0}\,
P^1_{3/2}\left(y_0\right)
\right)\left(\frac{\Delta}{\epsilon_F}\right)^{1/2}
\end{eqnarray}
where $y_0=-x_0 (x_0^2+1)^{-1/2}$. 
Note that, in this case ,$\Delta/\epsilon_F$ can be eliminated from these 
results to leave a single equation for $x_0=\mu/\Delta$ in terms of $p_Fa_0$. 

We now turn to the full evolution equations. At the present level of 
truncation, all of these can be obtained from the evolution of the 
effective potential. It is convenient to write this as $U(\rho,\mu,k)$
where $\rho=\phi^\dagger\phi$. The density and $Z_\phi$ are then given by
\begin{equation}
n=-\left.\frac{\partial U}{\partial\mu}\right|_{\rho=\Delta^2},\qquad
Z_\phi=-\,\frac{1}{2}\left.\frac{\partial^2 U}{\partial\rho\partial\mu}
\right|_{\rho=\Delta^2}.
\end{equation}

By evaluating the loop integrals in Eq.~(\ref{eq:Gamevol}) for a uniform 
boson field, we get evolution equation for $U$ at constant $\mu$,
\begin{eqnarray}
\partial_k U
&=&-\,\frac{1}{Z_\psi}\int\frac{d^3{\vec q}}{(2\pi)^3}\,\frac{E_{FR}}
{\sqrt{E_{FR}^2+\rho}}\,\sign\bigl(\epsilon(q)-\mu\bigr)\,\partial_k
R_F\nonumber\\
\noalign{\vskip 5pt}
&&+\,\frac{1}{2Z_\phi}\int\frac{d^3{\vec q}}{(2\pi)^3}\,
\frac{E_{BR}}{\sqrt{E_{BR}^2-V_B^2}}
\,\partial_kR_B,\label{eq:potevol}
\end{eqnarray}
where
\begin{equation}
E_{BR}(q)=\frac{Z_m}{2m}\,q^2+u_1+u_2(2\rho-\Delta^2)+R_B(q,k),\qquad
V_B=u_2\rho,
\end{equation}
and $E_{FR}$ is defined above.
Substituting our expansion on the left-hand side leads to a set of equations 
for the $u_n$, each of which is coupled
to the coefficient of the next term through the running of $\Delta^2$,
the position of the minimum of $U$. These equations have the form
\begin{equation}
\partial_k u_n-u_{n+1}\,\partial_k\Delta^2
=\left.\frac{\partial^n}{\partial \rho^n}
\Bigl(\partial_k U\Bigr)\right|_{\rho=\Delta^2}.
\end{equation}
In the symmetric phase ($\Delta^2=0$) they decouple. In the broken phase
they do not and, for example, the left-hand-side of the equation for 
$\partial_ku_2$ contains the coefficent $u_3$. 
We could simply set $u_3=0$, but a better
approximation can be obtained by substituting the form for $u_3(k)$ 
from the evolution with fermion loops only, as described above.
This is the approach we adopt here for $u_3$ and similar coefficients  
in the equations for $n$ and $Z_\phi$. It has the advantage of providing an 
approximation that becomes exact in situations where boson loops can
be neglected. 

In order to follow the evolution at constant density we allow $\mu$ to run 
with $k$ and define the total derivative
\begin{equation}
d_k=\partial_k+(d_k\mu)\,\frac{\partial}{\partial\mu}, 
\end{equation}
where $d_k\mu=d\mu/dk$. Applying this to $\partial U/\partial \mu$  
and demanding that $n$ is constant ($d_k n=0$) gives
\begin{equation}
-2Z_\phi\,d_k\Delta^2+\chi\,d_k\mu
=-\left.\frac{\partial}{\partial \mu}
\Bigl(\partial_k U\Bigr)\right|_{\rho=\Delta^2},
\label{eq:muevol}
\end{equation}
where we have introduced the fermion-number susceptibility
$\chi={\partial^2 U}/{\partial \mu^2}|_{\rho=\Delta^2}$.
The equations for the coefficients $u_n$ in the potential and $Z_\phi$ are
\begin{eqnarray}
d_k u_0+n\,d_k\mu
&=&\left.\partial_k U\right|_{\rho=\Delta^2},\\
\noalign{\vskip 5pt}
d_ku_1-u_2\,d_k\Delta^2+2Z_\phi\,d_k\mu
&=&\left.\frac{\partial}{\partial \rho}
\Bigl(\partial_k U\Bigr)\right|_{\rho=\Delta^2},\\
\noalign{\vskip 5pt}
d_k u_2-u_3\,d_k\Delta^2+2z_{\phi 1}\,d_k\mu
&=&\left.\frac{\partial^2}{\partial \rho^2}
\Bigl(\partial_k U\Bigr)\right|_{\rho=\Delta^2},\\
\noalign{\vskip 5pt}
d_k Z_\phi-z_{\phi 1}\,d_k\Delta^2+\frac{1}{2}\,\chi'\,d_k\mu
&=&-\,\frac{1}{2}\left.\frac{\partial^2}{\partial \mu\partial\rho}
\Bigl(\partial_k U\Bigr)\right|_{\rho=\Delta^2},
\end{eqnarray}
where we have defined\begin{equation}
z_{\phi 1}=-\,\frac{1}{2}\left.\frac{\partial^3 U}{\partial^2\rho\partial\mu}
\right|_{\rho=\Delta^2}\qquad
\chi'=\left.\frac{\partial^3 U}{\partial \mu^2\partial\rho}
\right|_{\rho=\Delta^2}.
\end{equation}
The driving terms in these evolution equations are given by appropriate 
derivatives of Eq.~(\ref{eq:potevol}). In the symmetric phase we evaluate 
these expressions at $\Delta^2=0$. The driving term of Eq.~(\ref{eq:muevol}) 
vanishes in this case, and hence $\mu$ remains constant. In the broken phase 
we keep $\Delta^2$ non-zero and set $u_1=0$. 

On the left-hand sides of these equations, the coefficients $u_3$, $z_{\phi 1}$, 
$\chi$ and $\chi'$ all correspond to terms outside our present truncation and 
so we replace them by their expressions from evolution with fermion loops 
only. This approximation provides a closed set of
equations, whose solutions agree with the exact results when boson loops are
neglected. Underlying it is an assumption is that bosonic contributions to the
evolution are not too large  compared to fermionic ones. We shall investigate this
assumption below, and find that it seems to hold for a wide range of values
of $p_Fa_0$. In future work, we hope to be able to check it further by
including more bosonic terms in our ansatz for $\Gamma$ and examining 
convergence with respect to the number of terms.

We integrate the resulting differential equations numerically. 
In this work, we use regulator functions with the forms
\begin{equation}
R_{F}(q,k;p_{F},\sigma)  =  \frac{k^{2}}{2M}\,\theta(q-p_{F},k;\sigma),\qquad
R_{B}(q,k;\sigma)  =  \frac{k^{2}}{2m}\,\theta(q,k;\sigma).
\end{equation}
where $\theta$ is the smoothed step-function
\begin{equation}
\theta(q,k;\sigma)  = \frac{1}{2\erf(1/\sigma)} 
\left[\erf\left(-\frac{(q+k)}{k\sigma}\right)
+\erf\left(-\frac{(q-k)}{k\sigma}\right)\right],
\end{equation}
with $\sigma$ being a parameter controlling the sharpness of the step.
The derivative $\partial_k\theta$ falls off for $k\rightarrow\infty$ and so 
$\partial_kR_{B,F}$ provides a UV cut-off on the loop integrals in
Eq.~(\ref{eq:potevol}) \cite{BTW02}.

We run the evolution down towards $k=0$ starting from some large, fixed
scale, $k=K$. The initial conditions for this are obtained by matching 
onto the standard evolution in vacuum for $k\ge K$ using
Eq.~(\ref{eq:UF}) and related expressions.
Differentiating $U^{\text{MF}}$ with respect to $\Delta^2$ at $\Delta=0$ 
we get the initial condition on $u_1$,
\begin{equation}
u_1(K)=-\frac{M}{4\pi a_0}+\frac{1}{2}\,\int\dtn{q} \left(
\frac{\sign(\epsilon(q)-\mu)}{E_{FR}(q,K)}-\frac{1}{\epsilon(q)}\right).
\label{eq:u1Kfull}
\end{equation}
The second term in the integral contains exactly the same linearly
divergent term as in the free inverse $T$ matrix. It cancels with 
the similar divergence in the first term to leave a finite result.
This result is linear in $K$, reflecting the divergence of the
underlying integral. It also contains a correction for the fact that 
our regulator depends on the density (through $p_F$). The initial
values of $u_2$ and $Z_\phi$ can be obtained similarly
from second derivatives of $U^{\text{MF}}$.

Although our results can be applied to many systems, 
for definiteness we concentrate initially on parameter values relevant to 
applications to neutron matter: $M=4.76$~fm$^{-1}$, $p_F=1.37$~fm$^{-1}$, 
and large two-body scattering lengths ($|a_0|\gsim 1$~fm). 
We then explore a wider range of values for $p_F a_0$. 
We have checked for dependence of our results on the starting 
scale $K$ and find that this is undetectable as long as $K>5$~fm$^{-1}$ 
(about $4p_F$). Similarly, we get numerically indistinguishable results for 
a range of values of the width parameter $\sigma$. 

\begin{figure}
\begin{center}
\includegraphics[width=11cm,  keepaspectratio,clip]{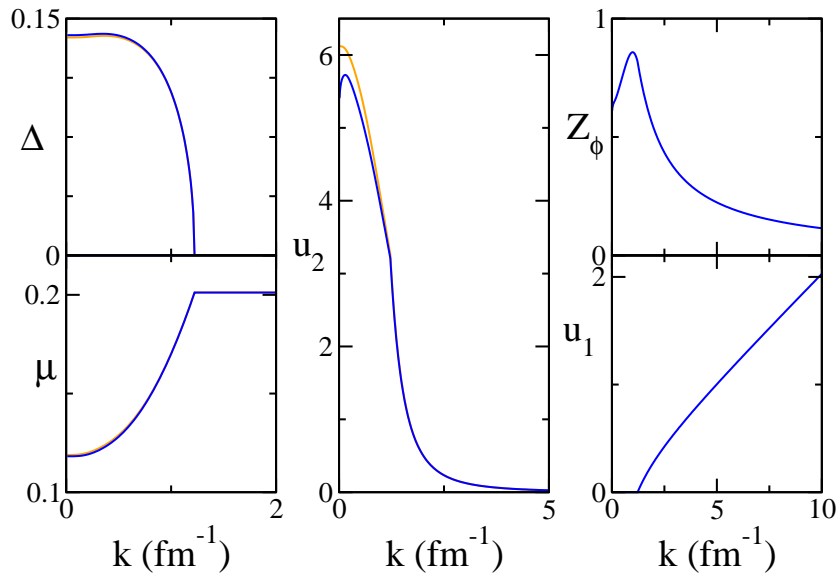}
\end{center}
\caption{\label{fig:running}Numerical solutions to the evolution equations
for infinite $a_0$ and $p_F=1.37$~fm, starting from $K=16$~fm$^{-1}$.
We show the evolution of all relevant 
parameters for the cases of fermion loops only (orange/grey lines), and  
of bosonic loops with a running $Z_\phi$ (blue/black lines). All 
quantities are expressed in appropriate powers of fm$^{-1}$.}
\end{figure}

Some typical solutions for the evolution equations are given in 
Fig.~\ref{fig:running}, for the case of infinite $a_0$. We compare  
two different approximation schemes, one where we have fermion loops only, 
and one where we include boson loops as well and we allow $Z_\phi$ to run.
For large values of $k$ the system remains in the symmetric phase. 
At $k_{\text{crit}}\simeq 1.2$~fm$^{-1}$, $u_1$ vanishes and below that 
point the system is in the broken phase where we plot $\Delta$ instead of $u_1$. 

One immediate observation is that the contributions of boson loops are small. 
Indeed in the symmetric phase their inclusion has essentially no effect.
Below the transition they do become visible, particularly in $u_2$.
However their effects on the gap are even smaller, at most $\sim 1\%$ 
for $1/|p_Fa_0|<1$. In this region, the main effect of the bosons is a small
enhancement of the gap related to the reduction in $u_2$ as $k\rightarrow 0$.
Not plotted are results including boson loops but with $Z_\phi$ fixed at unity;
the results are vitually indistinguishable from those with running $Z_\phi$.
The same is true if we set $Z_m=Z_\phi$ rather than $Z_m=1$.

\begin{figure}
\begin{center}
\includegraphics[width=9cm,keepaspectratio,clip]{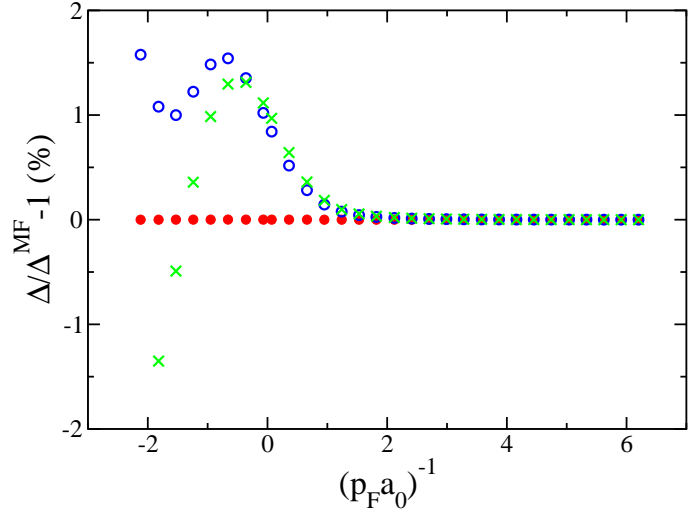}
\end{center}
\caption{\label{fig:ratio} Fractional deviations (in \%) 
of the gaps for various numerical solutions
from the analytic mean-field result. All cases are for
$p_F=1.37 \text{fm}^{-1}$. Solid dots (red) denote the results from a
calculation with fermion loops only, the open circles (blue) show the
results when boson loops are added, and the crosses (green) show the
effect of allowing $Z_{\phi}$ to run as well.}
\end{figure}

It is also instructive to examine the behaviour of the gap over a wider range
of $1/(p_{F}a_{0})$. The overall picture the same as found in Ref.~\cite{Babaev}
for fermion loops only, with a crossover from BCS pairing (with positive 
$\mu\simeq \epsilon_F$) for $1/(p_{F}a_{0})<0$ to BEC (with increasingly 
negative values for $\mu$) for $1/(p_{F}a_{0})>0$. For negative 
$1/(p_{F}a_{0})$, the gap lies close to the exponential 
curve of the weak coupling limit, Eq.~(\ref{eq:lowd}). However a closer look
at these results, as in Fig.~\ref{fig:ratio}, shows that deviations from 
mean field behaviour are present in this region and become increasingly
noticable for weaker couplings or lower densities.

We are not able to follow our results beyond $1/(p_{F}a_{0})\sim -2$.
This is because of the non-analyticity of the effective action for small 
gaps which means that our expansion Eq.~(\ref{eq:potexp}) is no longer 
adequate. For example, the fermion loops contain a term
$\phi^\dagger\phi\log(\phi^\dagger\phi)$ which gives a divergent 
contribution to $u_2$. Although the complete fermionic part has been 
incorporated in our equations, the analogous boson effects are not included. 
Some of our results suggest that the gap vanishes as a power rather than an 
exponential, but a definite conclusion will have to await a more complete 
calculation.

The dominance of fermion loops for most values of $1/(p_{F}a_{0})$
may seem surprising. However this is the region of BEC
or the crossover to it, where the mean boson field (the gap) is large. 
In fact the gap is particularly insensitive to boson fluctuations  as
a result of cancellations between the direct contribution to
the running of $\Delta^2$ and indirect ones via $u_2$.
Contributions to other parameters such as $u_2$ and $Z_\phi$ are larger, 
$\sim 10\%$ for $ 1/(p_{F}a_{0})\simeq 0$. Our results imply that 
these effects are likely to be important for neutron matter, where
calculations with realistic interactions lead to gaps of at most 5~MeV 
($\sim 0.1\epsilon_F$) \cite{DHJ03}.

However, if we were to apply our current calculation to neutron matter, 
we would find a gap comparable to $\epsilon_F$, of the order of 30~MeV.
Fayans has given a simple explanation for the smaller values found 
using more sophisticated treatments \cite{Fayans}.\footnote{A similar result 
can be obtained using the results of Ref.~\cite{Khodel} or \cite{EHJ98}.
Note that the derivation of the effective range formula in the second 
reference is unfortunately incorrect, due to the use of a misleading result 
quoted in Ref.~\cite{CS89}.} The argument can be given most succinctly 
for weak coupling, where the gap satisfies a generalisation of
Eq.~(\ref{eq:lowd}),
\begin{equation}
\Delta=({8}/{e^2})\epsilon_F \exp\left(-\,({\pi}/{2})
\cot\bigl(\delta(p_F)\bigr) \right).
\end{equation}
For nucleon-nucleon scattering, $\cot\delta$ increases relatively quickly
with momentum and the resulting reduction in the gap is substantial.
We therefore expect that an extension of our approach to include the 
effective range should capture this physics.

There a number of improvements which could be made to our approach. 
Adding an effective range is clearly an important one. 
Another is the evolution of the boson kinetic mass ($Z_m$) 
since this will allow a scaling analysis of boson loops for small gaps. 
We should also include running of the fermion renormalisation 
constants and ``Yukawa'' coupling, to provide a full treatment of the action 
(\ref{eq:Gansatz}). Beyond that we would like to treat explicitly the 
particle-hole channels (RPA phonons) since these contain important physics.
They will also allow us to remove the ``Fierz ambiguity'' associated with
our bosonisation of the underlying contact interaction \cite{JW03}.
We plan to explore these extensions in future studies.

This research was funded by the EPSRC.

\end{document}